\documentclass[aps,pra,superscriptaddress,twocolumn]{revtex4-2}
\usepackage{graphicx}
\usepackage{amsmath}
\usepackage{amssymb}
\usepackage{booktabs}
\usepackage[colorlinks=true, allcolors=blue]{hyperref}
\usepackage{bm}
\usepackage{textcomp}
\usepackage{siunitx}
\begin{document}

\title{A bright source of high-dimensional spatially entangled photon pairs based on beam-like Type-II SPDC}

\author{Ming Zhang}
\author{Zhiming Qing}
\affiliation{National Laboratory of Solid State Microstructures and School of Physics, Nanjing University, Nanjing 210093, China}
\affiliation{Collaborative Innovation Center of Advanced Microstructures, Nanjing University, Nanjing 210093, China}

\author{Zhi-Cheng Ren}
\email{zcren@nju.edu.cn}
\affiliation{National Laboratory of Solid State Microstructures and School of Physics, Nanjing University, Nanjing 210093, China}
\affiliation{Collaborative Innovation Center of Advanced Microstructures, Nanjing University, Nanjing 210093, China}
\affiliation{Collaborative Innovation Center of Extreme Optics, Shanxi University, Tiyuan 030006, China}

\author{Jianping Ding}
\affiliation{National Laboratory of Solid State Microstructures and School of Physics, Nanjing University, Nanjing 210093, China}
\affiliation{Collaborative Innovation Center of Advanced Microstructures, Nanjing University, Nanjing 210093, China}

\author{Xi-Lin Wang}
\email{xilinwang@nju.edu.cn}
\affiliation{National Laboratory of Solid State Microstructures and School of Physics, Nanjing University, Nanjing 210093, China}
\affiliation{Collaborative Innovation Center of Advanced Microstructures, Nanjing University, Nanjing 210093, China}
\affiliation{Jiangsu Physical Science Research Center, Nanjing 210093, China}
\affiliation{Jiangsu Key Laboratory of Quantum Information Science and Technology, Nanjing University, Nanjing 210093, China}
\affiliation{Hefei National Laboratory, Hefei 230088, China}
\affiliation{Synergetic Innovation Center of Quantum Information and Quantum Physics, University of Science and Technology of China, Hefei 230026, China}

\author{Hui-Tian Wang}
\affiliation{National Laboratory of Solid State Microstructures and School of Physics, Nanjing University, Nanjing 210093, China}
\affiliation{Collaborative Innovation Center of Advanced Microstructures, Nanjing University, Nanjing 210093, China}
\affiliation{Collaborative Innovation Center of Extreme Optics, Shanxi University, Tiyuan 030006, China}

\begin{abstract}
Bright sources of high-dimensional spatially entangled photon pairs are increasingly required for the rapid advancement of quantum technologies. However, simultaneously achieving high brightness and high-dimensional spatial entanglement through spontaneous parametric down-conversion (SPDC) remains challenging. Here, we demonstrate a bright source of high-dimensional spatially entangled photon pairs using beam-like Type-II SPDC in a $\beta$-barium borate crystal. Compared with a conventional collinear Type-I source, it exhibits a $\sim$1.56-fold enhancement in the detected single-photon count rate and a $\sim$1.45-fold enhancement in the coincidence count rate. Measurements of position correlations and momentum anti-correlations certify at least 62-dimensional spatial entanglement within a 121-dimensional basis. These correlations are further exploited to demonstrate quantum imaging in both the position and momentum domains. These characteristics make the source a practical resource for high-dimensional photonic quantum technologies.
\end{abstract}

\maketitle

\section{Introduction}
\label{sec:intro}

Quantum entanglement is a central resource for quantum information science~\cite{Horodecki2009, Erhard2020, Zhang2024}. In particular, high-dimensional photonic entanglement offers increased information capacity~\cite{Cozzolino2019} and enhanced resilience to noise~\cite{Ecker2019}, providing advantages for quantum communication~\cite{Cozzolino2019, Luo2019}, quantum imaging~\cite{Moreau2019, Defienne2024}, quantum metrology~\cite{Ciampini2016}, and quantum information processing~\cite{Paesani2021, Chi2022, Chi2023}. The transverse position and momentum of photon pairs are especially attractive for high-dimensional encoding because they provide access to a large spatial Hilbert space~\cite{Law2004,Howell2004}. Spontaneous parametric down-conversion (SPDC) in nonlinear crystals is one of the most established methods for generating such spatially entangled photon pairs~\cite{Walborn2010,Schneeloch2016}. However, practical applications  require sources that combine high-dimensional entanglement with high detected photon-pair rates and efficient photon collection. Conventional bulk Type-I and Type-II SPDC sources typically exhibit cone-like emission, such that only a limited fraction of the down-converted photon pairs can be coupled into selected collection modes, thereby restricting the collected photon-pair brightness~\cite{Kwon2008, Zhang2021}. Beam-like Type-II phase matching provides an attractive alternative, producing two separated, orthogonally polarized output beams that permit efficient collection and deterministic separation of the signal and idler photons~\cite{Takeuchi2001, Niu2008, Wang2016, Liu2026}. Combining these properties with high-dimensional spatial entanglement is therefore desirable for applications requiring high photon brightness, such as quantum imaging and high-dimensional quantum-state characterization~\cite{Edgar2012, Ndagano2020, Defienne2022}.

Here, we demonstrate a bright source of spatially entangled photon pairs based on beam-like Type-II SPDC in a $\beta$-barium borate (BBO) crystal. Compared with a conventional collinear Type-I SPDC source under identical pumping and acquisition conditions, the beam-like Type-II source exhibits a $\sim$1.56-fold enhancement in the single-photon count rate and a $\sim$1.45-fold enhancement in the biphoton coincidence count rate. The orthogonal polarizations of the Type-II photon pairs additionally enable deterministic separation of the signal and idler photons. Using this high-brightness source, we perform two representative proof-of-principle experiments. First, spatial correlations are measured in the mutually unbiased position and momentum bases, certifying a spatial entanglement dimensionality of 62, thus demonstrating the capability of the source for high-dimensional quantum information processing. Second, the position correlations and momentum anti-correlations of the biphoton state are exploited to realize correlation imaging in the position and momentum domains, respectively. These results indicate that the beam-like Type-II SPDC source is a versatile platform for generating bright, high-dimensional spatial entanglement, with potential applications in quantum imaging, quantum communication, and other scalable photonic quantum technologies.

\section{Theory}
\label{sec:theory}
In our configuration, photon pairs are generated via SPDC in a BBO crystal operated under a beam-like Type-II phase-matching condition. The biphoton state at the crystal output plane can be written as~\cite{Monken1998}

\begin{equation}
\label{eq:biphoton_state}
|\Psi\rangle =
\int d^2\mathbf{r}_s\,d^2\mathbf{r}_i\,
\Psi(\mathbf{r}_s,\mathbf{r}_i) \hat{a}_{s,V}^{\dagger}(\mathbf{r}_s) \hat{a}_{i,H}^{\dagger}(\mathbf{r}_i) |0\rangle,
\end{equation}

where $\mathbf{r}_s$ and $\mathbf{r}_i$ denote the transverse positions of the signal and idler photons, respectively. The operator $\hat{a}_{k,\mu}^{\dagger}(\mathbf{r}_k)$ creates a photon with polarization $\mu$ at $\mathbf{r}_k$ in path $k$, where $k=s$ ($k=i$) denotes the signal (idler) path and $\mu=H$ ($\mu=V$) denotes horizontal (vertical) polarization. Under the paraxial and double-Gaussian approximations, the biphoton wavefunction generated by a Gaussian pump beam with waist $w$ can be written as~\cite{Law2004, Walborn2010, Schneeloch2016}
\begin{equation}
\label{eq:psi}
\Psi(\mathbf{r}_s,\mathbf{r}_i) = A\exp\left(-\frac{|\mathbf{r}_{-}|^2}{4\sigma_{r-}^2}\right)
\exp\left(-\frac{|\mathbf{r}_{+}|^2}{4\sigma_{r+}^2}\right),
\end{equation}
where $A$ is a normalization constant and
$\mathbf{r}_{+}\equiv\mathbf{r}_s+\mathbf{r}_i$ and $\mathbf{r}_{-}\equiv\mathbf{r}_s-\mathbf{r}_i$ are the sum and minus transverse position coordinates, respectively. Their spatial  widths are
\begin{equation}
\sigma_{r+}=2w,
\qquad
\sigma_{r-}=\sqrt{\frac{2\lambda_p L}{9\pi}},
\end{equation}
where $\lambda_p$ is the pump wavelength and $L$ is the crystal length. Here, $\sigma_{r-}$ represents the position correlation width.

The corresponding momentum-space representation can be obtained by a Fourier transform of the position-space wavefunction Eq. (\ref{eq:psi}), yielding
\begin{equation}
\label{eq:fourier}
\Phi(\mathbf{q}_s,\mathbf{q}_i) = A' \exp\left(-\frac{|\mathbf{q}_{+}|^2}{4\sigma_{p+}^2}\right)
\exp\left(-\frac{|\mathbf{q}_{-}|^2}{4\sigma_{p-}^2}\right),
\end{equation}
where $A'$ is a normalization constant and
$\mathbf{q}_{\pm}\equiv\mathbf{q}_s \pm \mathbf{q}_i$. Here, $\mathbf{q}_s$ and $\mathbf{q}_i$ denote the Fourier-plane coordinates associated with the transverse momenta of the signal and idler photons. In the coordinate convention adopted here, the corresponding momentum widths are
\begin{equation}
\sigma_{p+}=\frac{2\lambda_p F}{\sigma_{r+}},
\qquad
\sigma_{p-}=\frac{2\lambda_p F}{\sigma_{r-}},
\end{equation}
where $F$ denotes the effective focal length of the Fourier-transform system linking the crystal output plane and the imaging plane in the momentum domain. The parameter $\sigma_{p+}$ represents the momentum anticorrelation width.

In our configuration, the Type-II BBO crystal is pumped by a Gaussian beam with a sufficiently large beam waist, resulting in two Gaussian envelopes in the position domain satisfying that $\sigma_{r+}/\sigma_{r-}\gg1$, and therefore the narrow minus-coordinate distribution is concentrated around $\mathbf{r}_{-}=0$, corresponding to strong position correlations, $\mathbf{r}_s\approx\mathbf{r}_i$. The Fourier transform interchanges the relative widths of the two Gaussian envelopes. Consequently, $\sigma_{p-}/\sigma_{p+}\gg1$, and the narrow sum-coordinate distribution is concentrated around $\mathbf{q}_{+}=0$, yielding strong EPR-like momentum anti-correlations, $\mathbf{q}_s\approx-\mathbf{q}_i$.

The position domain joint probability distribution (JPD) is given by
\begin{equation}
\label{eq:jpd}
\Gamma(\mathbf{r}_s,\mathbf{r}_i)
=
\left|
\Psi(\mathbf{r}_s,\mathbf{r}_i)
\right|^2.
\end{equation}
Integrating the JPD over $\mathbf{r}_{+}$ [$\mathbf{r}_{-}$] yields the marginal distribution of sum-coordinate $\Gamma_+(\mathbf{r_+})$ [minus-coordinate $\Gamma_-(\mathbf{r_-})$]. These marginal distributions correspond to projections of the JPD onto the sum- and minus-coordinate, respectively.

Experimentally, the JPD is reconstructed from frame-by-frame coincidence measurements. For a dataset containing $M$ photon-counting frames, an estimator of the correlation signal is obtained by subtracting the accidental coincidences evaluated from consecutive frames from the coincidences measured within the same frame as~\cite{Defienne2022, Defienne2018}
\begin{equation}
\label{eq:coincidence}
\Gamma_{\mathrm{exp}}(\mathbf{r}_s,\mathbf{r}_i) = \frac{1}{M-1}\sum_{\ell=1}^{M-1}\left[I_{\ell}(\mathbf{r}_s)I_{\ell}(\mathbf{r}_i)-I_{\ell}(\mathbf{r}_s)I_{\ell+1}(\mathbf{r}_i)\right].
\end{equation}
Here, $I_{\ell}(\mathbf{r}_k)$, with $k\in\{s,i\}$, denotes the photon count recorded at $\mathbf{r}_k$ in the $\ell$-th frame of the signal ($k=s$) or idler ($k=i$) channel. The first term contains both photon-pair and accidental coincidences within the same acquisition frame. The second term is a time-shifted correlation constructed from adjacent frames and provides an estimate of the accidental coincidence events arising from statistically uncorrelated detections, including uncorrelated background photons, detector dark counts, and other random noise processes. Subtracting the second term from the first therefore isolates correlated photon-pair events from the accidental coincidence background. After normalization, $\Gamma_{\mathrm{exp}}(\mathbf{r}_s,\mathbf{r}_i)$ provides an experimental estimate of the biphoton JPD.

\section{Experimental Setup}
\label{sec:setup}
\begin{figure*}[htbp]
\centering
\includegraphics[width=0.95\linewidth]{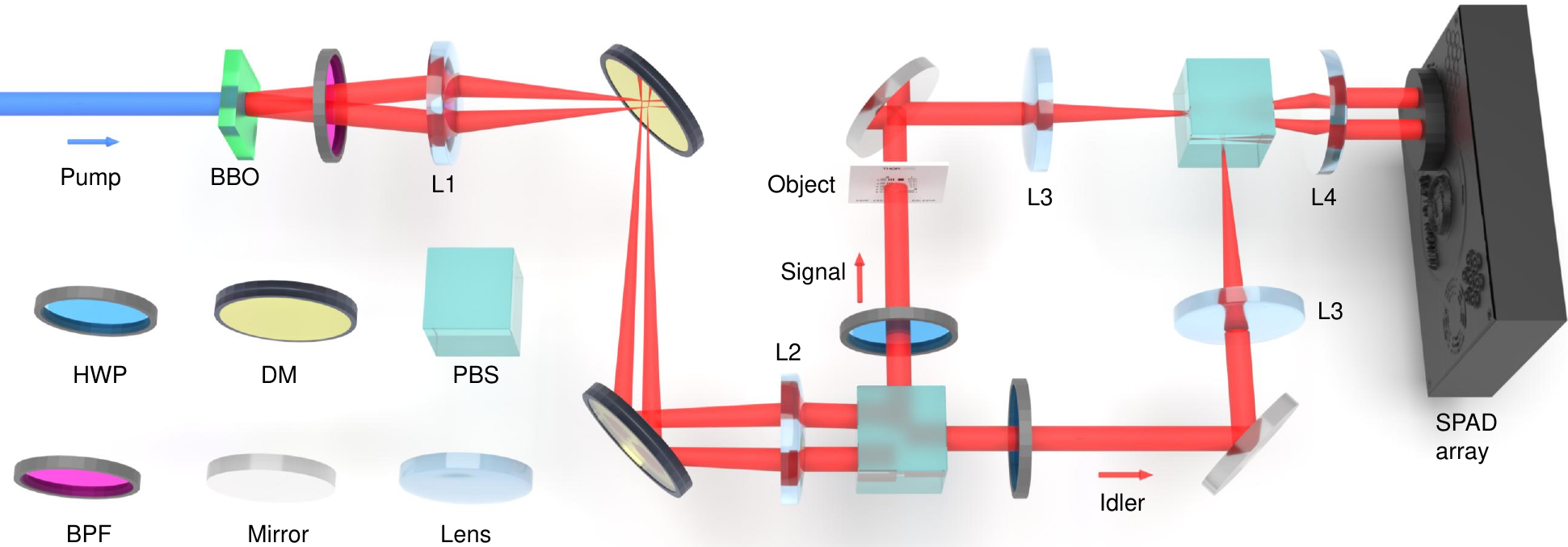}
\caption{Experimental setup. A 390-nm continuous-wave laser pumps a 2-mm-long BBO crystal operated under the beam-like Type-II phase-matching condition, to generate noncollinear signal and idler photon pairs via the SPDC. A bandpass filter (BPF) centered at 780~nm with an 8-nm full width at half maximum bandwidth selects frequency-degenerate photon pairs, while two dichroic mirrors (DM) at $45^{\circ}$ incidence reject the residual pump light. A polarizing beam splitter (PBS) separates the orthogonally polarized photons into the signal and idler paths. After polarization rotation by half-wave plates (HWP) oriented at $45^{\circ}$, the two paths are recombined at a second PBS and imaged onto two spatially separated regions of a single-photon avalanche diode (SPAD) array camera, thereby minimizing the contribution of local detector crosstalk to interchannel coincidences. For correlation-imaging measurements, an object is placed on the image plane of the first 4-$f$ system in the signal path. The two cascaded 4-$f$ systems are composed of lenses $L_1$--$L_4$ with focal lengths $f_1=75~\mathrm{mm}$, $f_2=200~\mathrm{mm}$, $f_3=100~\mathrm{mm}$, and $f_4=100~\mathrm{mm}$ or $50~\mathrm{mm}$. With $f_4=100~\mathrm{mm}$, the crystal output plane is imaged onto the camera for position-domain measurements; replacing $L_4$ with a 50-mm-focal-length lens images the crystal Fourier plane for momentum-domain measurements.}
\label{fig:setup}
\end{figure*}
The experimental setup is shown in Fig.~\ref{fig:setup}. A 390~nm continuous-wave laser with an output power of 120~mW pumped a 2-mm-long BBO crystal operated under beam-like Type-II phase-matching conditions, with cut angles of $\theta=42.62^{\circ}$ and $\varphi=30^{\circ}$. The pump beam with a beam waist of $\sim$164.49~\textmu m was normally incident on the crystal to generate signal and idler photon pairs via the SPDC. A bandpass filter (BPF) centered at 780~nm with an 8-nm full width at half maximum (FWHM) bandwidth was used to select frequency-degenerate signal and idler photon pairs, while a dichroic mirror (DM) further suppressed the residual pump light.

The photon pairs were detected by a SPAD512 array camera comprising $512\times512$ pixels with a pixel pitch of $16.38~\text{\textmu m}$ and a photon-detection efficiency of $\sim$17\% at 780~nm. To suppress local interpixel crosstalk from producing false photon-pair coincidences, the orthogonally polarized signal and idler photons were separated by a polarizing beam splitter (PBS), routed through distinct optical paths, and directed onto two spatially separated, nonoverlapping regions of the same sensor.

For position correlation measurements, two cascaded 4-$f$ imaging systems comprising the lens pairs $(L_1, L_2)$ and $(L_3, L_4)$ with corresponding focal lengths $(f_1, f_2)$ and $(f_3, f_4)$, were used to image the crystal output plane onto the SPAD camera, enabling direct measurement of the position-domain biphoton JPD. For momentum anti-correlation measurements, the lens $L_4$ in the second 4-$f$ system was replaced by a lens with its focal length $f_4/2$. With the remaining optical geometry unchanged, this replacement converts the final imaging stage from an image-plane configuration to a Fourier-transform configuration at the camera plane, enabling direct measurement of the biphoton JPD in the transverse-momentum domain. For correlation-imaging measurements, an object was placed at the image plane of the first 4-$f$ imaging system in the signal path.

\section{Characterization of the beam-like Type-II SPDC source}
\label{sec:performance}

\subsection{Characterization of Brightness}

\begin{figure}[htb]
\centering
\includegraphics[width=\linewidth]{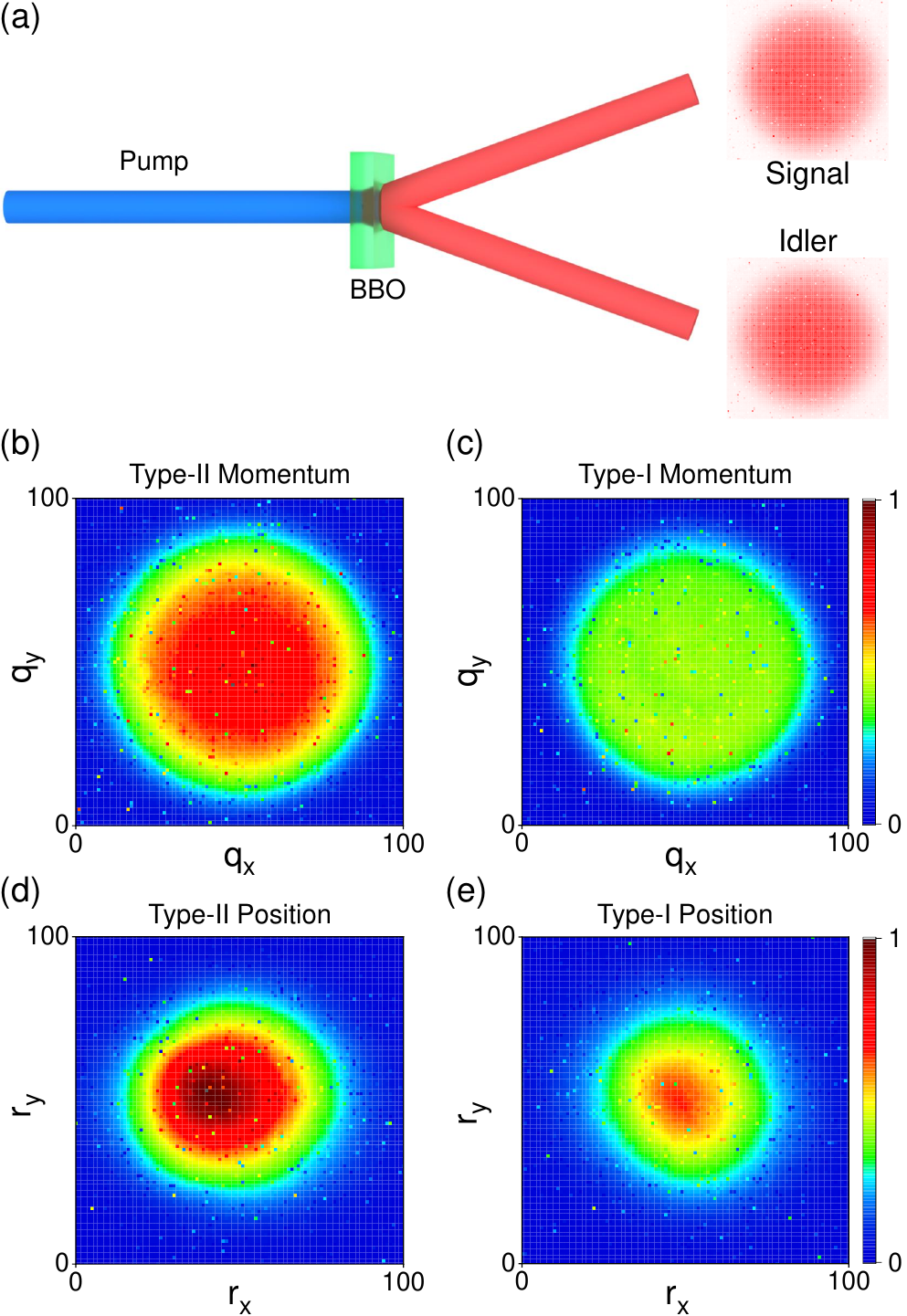}
\caption{\textbf{Characterization of the beam-like Type-II SPDC source.}
(a) Beam-like Type-II SPDC emission geometry and corresponding far-field photon-count distribution.
(b),(c) Measured momentum-domain single-photon count distributions for the beam-like Type-II and conventional collinear Type-I SPDC sources, respectively.
(d,e) Position-domain single-photon count distributions for the beam-like Type-II and Type-I SPDC sources, respectively.}
\label{fig:angular}
\end{figure}

The beam-like Type-II SPDC source generates noncollinear signal and idler photon pairs with an angular separation of $\sim$$3^{\circ}$, as illustrated in Fig.~\ref{fig:angular}(a). The corresponding far-field distribution exhibits two distinct emission spots associated with the signal and idler photons. To evaluate the source brightness, the beam-like Type-II source was compared with a conventional collinear Type-I SPDC source using a BBO crystal of the same length under identical pump and acquisition conditions. For both sources, the pump power was fixed at 120~mW. The generated photons were detected using the SPAD camera, which recorded $10^6$ frames with an exposure time of 0.5~ms per frame.

The single-photon brightness of the two sources was measured in both the momentum and position domains. Figures~\ref{fig:angular}(b) and \ref{fig:angular}(c) show the measured single-photon count distributions in the momentum domain, whereas Figs.~\ref{fig:angular}(d) and \ref{fig:angular}(e) show the corresponding distributions in the position domain. For the quantitative brightness comparison, the momentum-domain distributions were integrated and normalized by the total integration time and pump power. The resulting single-photon count rates were $1.9166\times10^4~\mathrm{Hz/mW}$ for the Type-I source and $2.9818\times10^4~\mathrm{Hz/mW}$ for the beam-like Type-II source, indicating a $\sim$1.56-fold enhancement in the detected single-photon count rate for the beam-like Type-II configuration.

To further assess the photon-pair brightness, the coincidence count rates of the two sources were compared. The measured coincidence count rates were $389.32~\mathrm{Hz/mW}$ for the Type-I source and $1127.60~\mathrm{Hz/mW}$ for the beam-like Type-II source. For the Type-II source, the orthogonally polarized signal and idler photons were deterministically separated into two detection channels by the PBS. By contrast, the identically polarized photon pairs generated by the Type-I source were probabilistically divided between the two detection channels by a balanced beam splitter, such that only half of the generated pairs resulted in one photon in each channel and were therefore registered as coincidence events. Correcting for this 50\% separation probability gives an effective Type-I coincidence count rate of $778.64~\mathrm{Hz/mW}$. After this correction, the photon-pair count rate of the beam-like Type-II source was $\sim$1.45 times that of the conventional collinear Type-I source.

The count rates reported here represent the detected rates measured by the SPAD camera rather than the intrinsic photon-generation rates of the sources. Their absolute values are limited by the quantum efficiency of the SPAD camera and the overall collection and transmission efficiencies of the optical system. Higher detected count rates could, in principle, be achieved using optimized optical coupling and point single-photon detectors with higher quantum efficiencies. Nevertheless, because both sources were characterized using the same imaging system, detector, and acquisition parameters, the present measurements enable a direct comparison of their brightness.

Clearly, the enhancements in the single-photon and coincidence biphoton count rates demonstrate that the beam-like Type-II source provides higher single-photon and photon-pair brightness than its conventional Type-I counterpart under the same experimental conditions. In addition, the orthogonal polarizations of the Type-II photon pairs enable their deterministic separation, providing a practical advantage for subsequent spatial correlation and quantum imaging measurements.

\begin{figure}[t]
\centering
\includegraphics[width=\linewidth]{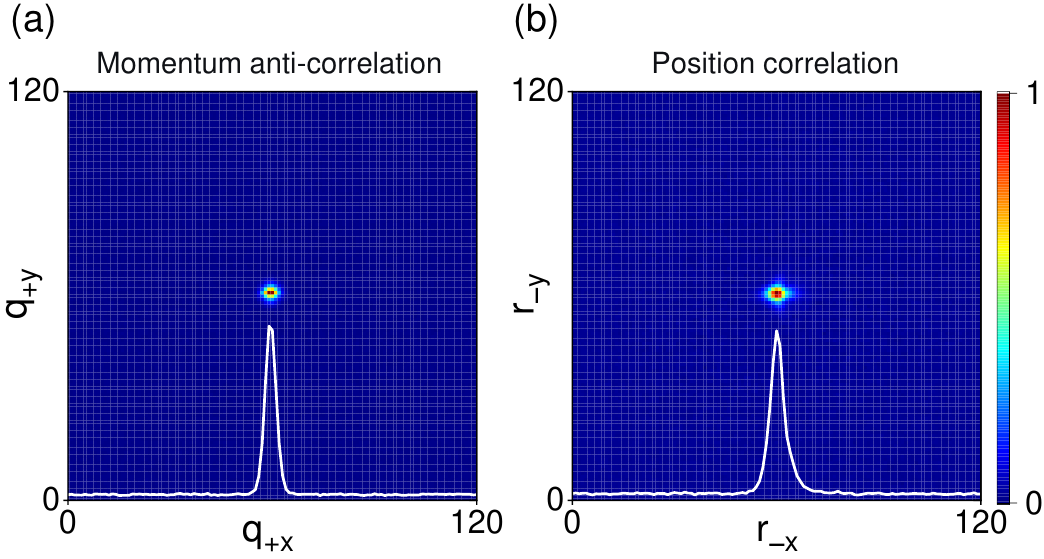}
\caption{\textbf{Characterization of biphoton spatial correlations.}
(a) Momentum anti-correlation distribution obtained by projecting the reconstructed momentum-domain JPD onto the sum-coordinate plane.
(b) Position correlation distribution obtained by projecting the reconstructed position-domain JPD onto the minus-coordinate plane.
The overlaid curves show horizontal cross sections through the maxima of the projected distributions together with the corresponding Gaussian fits used to extract the correlation widths.}
\label{fig:correlation}
\end{figure}

\subsection{Characterization of Spatial Correlation }

To characterize the spatial correlations of the generated biphoton state, we reconstructed the JPD in both the momentum and position domains. Within the double-Gaussian approximation introduced above, the projected correlation distributions are expected to exhibit Gaussian profiles. We therefore fitted the measured profiles with Gaussian functions to extract the experimental correlation widths.

For the momentum-domain measurement, the reconstructed biphoton JPD was projected onto the sum-coordinate plane, yielding the momentum anti-correlation distribution shown in Fig.~\ref{fig:correlation}(a). A one-dimensional profile passing through the maximum of the projected distribution was extracted and fitted with a Gaussian function. The effective focal length of the Fourier-transform system was $F=37.5~\mathrm{mm}$, corresponding to a theoretical momentum anti-correlation width of $\sigma_{p+}^{\mathrm{th}}=22.23~\text{\textmu m}$ at the SPAD camera plane. The Gaussian fitting yielded an experimental width of $\sigma_{p+}^{\mathrm{exp}}=23.75~\text{\textmu m}$, in good agreement with the theoretical prediction.

For the position-domain measurement, the reconstructed biphoton JPD was projected onto the minus-coordinate plane, as shown in Fig.~\ref{fig:correlation}(b). The position correlation width was then determined by fitting a Gaussian function to a one-dimensional cross section through the maximum of the projected distribution. The theoretical position correlation width at the crystal output plane was $\sigma_{r-}^{\mathrm{th}}=7.43~\text{\textmu m}$. Accounting for the overall imaging magnification of $m=f_2/f_1=8/3$, the predicted width at the SPAD camera plane was $m\sigma_{r-}^{\mathrm{th}}=19.81~\text{\textmu m}$. The Gaussian fitting to the measured distribution yielded an experimental width of $m\sigma_{r-}^{\mathrm{exp}}=26.54~\text{\textmu m}$.

The experimentally measured correlation widths are slightly broader than the corresponding theoretical values. This discrepancy may arise from residual spatial walk-off, as well as geometric aberrations and finite spatial sampling by the SPAD array. The finite sampling is particularly relevant because the predicted correlation widths are comparable to the camera pixel pitch of $16.38~\text{\textmu m}$.

\section{Certification of High-dimensional Spatial Entanglement}
\label{sec:dimension}

\begin{figure}[t]
\centering
\includegraphics[width=3.4in]{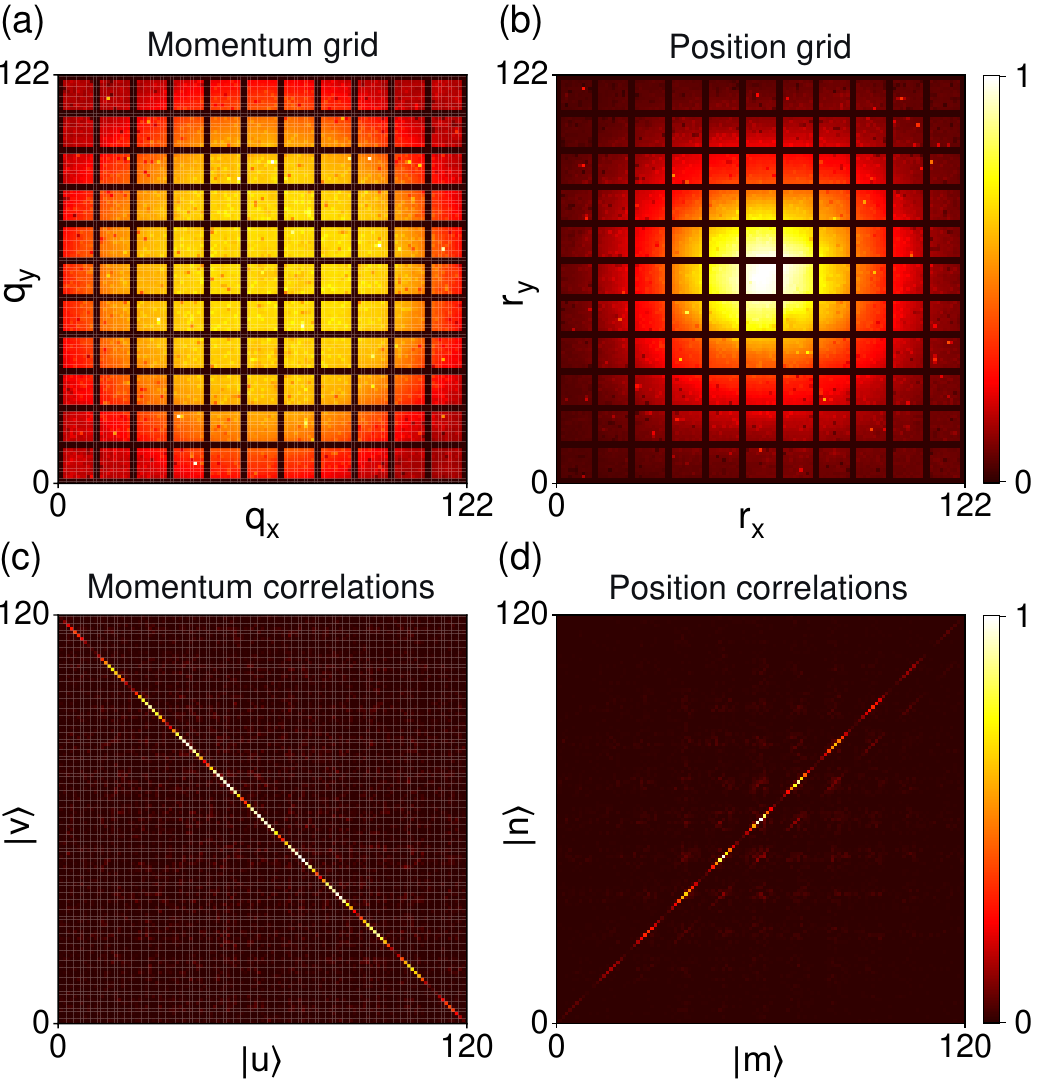}
\caption{\textbf{Certification of high-dimensional spatial entanglement.}
(a), (b) Spatial-mode discretization in the momentum and position domains, respectively, using uniform $11\times11$ macropixel grids corresponding to $d=121$ modes in each basis.
(c), (d) Measured correlation matrices between the 121 spatial modes in the momentum and position bases, respectively.}
\label{fig:dimension}
\end{figure}

Beyond characterizing the biphoton spatial correlations, we certify the dimensionality of the spatial entanglement generated by the beam-like Type-II SPDC source using correlation measurements in two mutually unbiased bases (MUBs), following the two-basis certification method of Refs.~\cite{Ndagano2020, Bavaresco2018, Valencia2020}. In the experiment, we choose discretized position and momentum bases, which are related by a discrete Fourier transform and therefore constitute a pair of MUBs.

For each photon, the transverse position and momentum spaces are each discretized into 121 modes using an $11\times11$ macropixel grids, corresponding to a Hilbert-space dimensionality of $ d=11^2=121 $. To suppress the crosstalk between neighboring modes, a gap of $32.76~\text{\textmu m}$ is introduced between two adjacent macropixels. The grid size and intermode spacing are jointly optimized to minimize crosstalk while preserving as many accessible spatial modes as possible, thereby maximizing the certifiable entanglement dimensionality. The resulting mode grids in the momentum and position domains are shown in Figs.~\ref{fig:dimension}(a) and \ref{fig:dimension}(b), respectively.

Following the two-basis certification protocol, the measured correlation matrices are used to determine a lower bound, $F_{\min}$, on the fidelity
\begin{equation}
F(\Phi_\rho,\Phi_d)=\mathrm{Tr}(\vert \Phi_d \rangle \langle \Phi_d \vert \Phi_\rho)  
\end{equation}
between the experimentally generated state $\vert \Phi_\rho \rangle$ and the $d$-dimensional maximally entangled target state
\begin{equation}
\vert \Phi_d \rangle=\frac{1}{\sqrt{d}} \sum_{m=0}^{d-1} \vert m, m \rangle .
\end{equation}

The detailed derivation of $F_{\min}$ is provided in the Appendix A. For the maximally entangled target state $\vert \Phi_d \rangle$, the fidelity of any stat $\vert \Phi_\rho \rangle$ with Schmidt number $k$ is bounded by
\begin{equation}
B_k(\Phi_d) = \frac{k}{d}.
\end{equation}
Therefore, a fidelity lower bound satisfying $F_{\min}>B_k(\Phi_d)$ certifies a spatial entanglement dimensionality of at least $k+1$.

The experimentally measured correlation matrices in the momentum and position bases are shown in Figs.~\ref{fig:dimension}(c) and \ref{fig:dimension}(d), respectively. The two-basis analysis yields
\begin{equation}
\label{eq:fidelity_dimension}
F_{\min}=0.5052>B_{61}=\frac{61}{121}=0.5041.
\end{equation}
Thus, the biphoton state generated by our beam-like Type-II SPDC source is certified to exhibit spatial entanglement in at least 62 dimensions.

\begin{figure}[t]
\centering
\includegraphics[width=\linewidth]{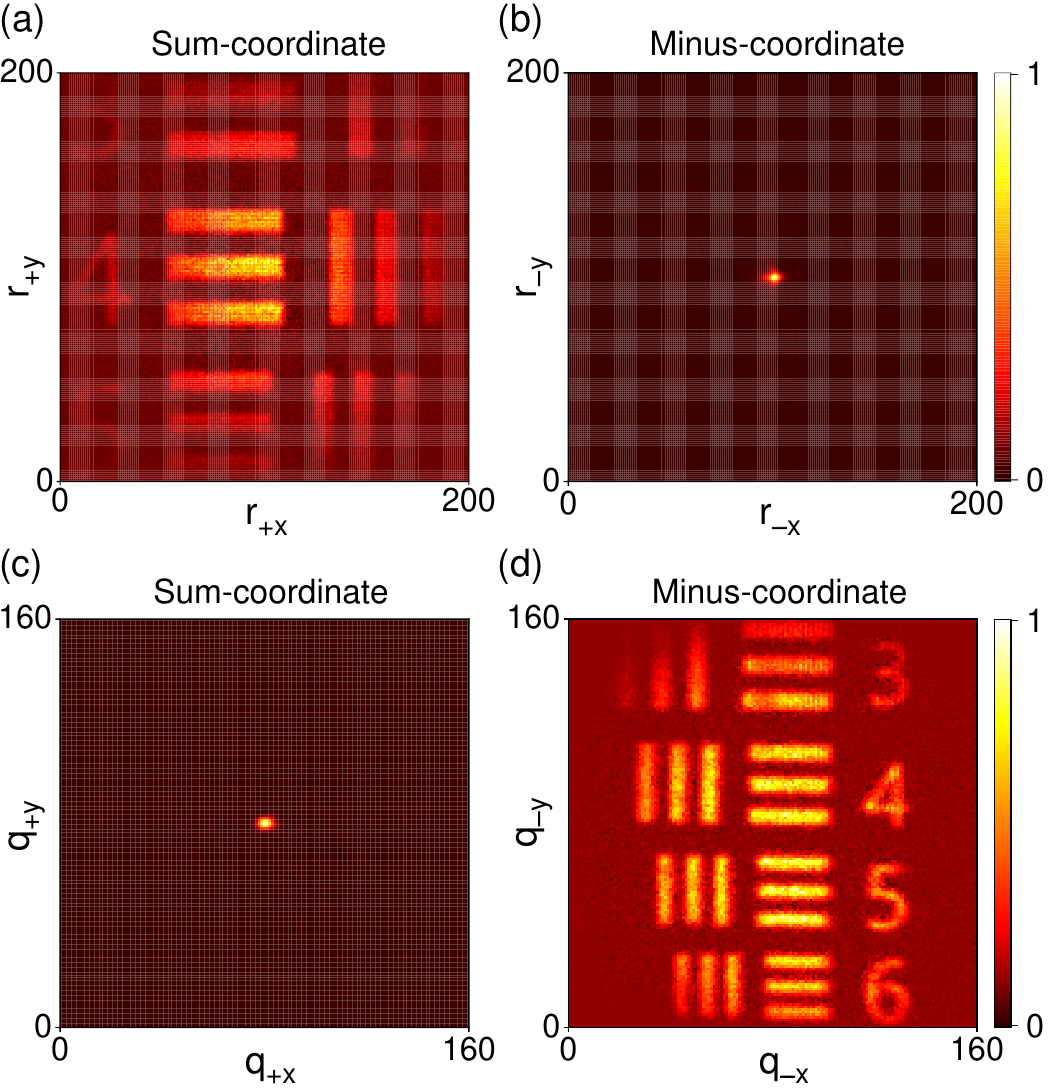}
\caption{\textbf{Correlation imaging using biphoton spatial correlations.}
(a) Position correlation image reconstructed by projecting the measured JPD onto the sum-coordinate.
(b) Corresponding Position correlation distribution obtained by projecting the JPD onto the minus-coordinate space.
(c) Momentum anti-correlation distribution obtained by projecting the Fourier-plane JPD onto the sum-coordinate.
(d) Object reconstructed by corresponding momentum anti-correlation imaging.}
\label{fig:imaging}
\end{figure}

\section{Correlation Imaging}
\label{sec:imaging}

The spatial correlations of the biphoton state generated by the beam-like Type-II SPDC source can be directly exploited for correlation imaging. 
For position correlation imaging, an object placed in the signal arm at the image plane of the first 4-$f$ imaging system modulates the biphoton state through its transmission function $T(\mathbf{r}_s)$, where $\mathbf{r}_s=(\mathbf{r}_++\mathbf{r}_-)/2$. Neglecting the magnification of the imaging system for simplicity, the biphoton wavefunction after modulation by an object in the signal arm is given by

\begin{equation}
\label{eq:modulated}
\Psi_T(\mathbf{r}_+,\mathbf{r}_-)=
A\,T\!\left(\frac{\mathbf{r}_+ + \mathbf{r}_-}{2}\right)\exp\!\left(\frac{|\mathbf{r}_+|^2}{4\sigma_{r+}^2}\right)\exp\!\left(-\frac{|\mathbf{r}_-|^2}{4\sigma_{r-}^2}\right).
\end{equation}

Under the condition $\sigma_{r+}/\sigma_{r-}\gg1$, the minus-coordinate envelope is much narrower than the sum-coordinate envelope and can therefore be approximated by a two-dimensional Dirac delta function, effectively acting as a narrow position correlation filter. The corresponding JPD is therefore expressed as

\begin{equation}
\label{eq:gmmat}
\Gamma_T(\mathbf{r}_+,\mathbf{r}_-)
\propto
\left|T\!\left(\frac{\mathbf{r}_+}{2}\right)\right|^2
\exp\!\left(-\frac{|\mathbf{r}_+|^2}{2\sigma_{r+}^2}\right)
\delta(\mathbf{r}_-).
\end{equation}

The delta function confines the coincidence events to $\mathbf{r}_-=0$, corresponding to $\mathbf{r}_s=\mathbf{r}_i$, and suppresses contributions from photon pairs with large transverse separations. Projection of the JPD onto the sum-coordinate then yields

\begin{equation}
\label{eq:imaging_position}
\begin{split}
P_+(\mathbf{r}_+)
&=\int d^2\mathbf{r}_-\, \Gamma_T(\mathbf{r}_+,\mathbf{r}_-) \\
&\propto\left|T\!\left(\frac{\mathbf{r}_+}{2}\right)\right|^2 \exp\!\left(-\frac{|\mathbf{r}_+|^2}{2\sigma_{r+}^2}\right).
\end{split}
\end{equation}

Thus, the reconstructed object is obtained from the projection onto the sum-coordinate space, whereas the complementary projection onto the minus-coordinate space exhibits a narrow position correlation peak, as shown in Figs.~\ref{fig:imaging}(a) and \ref{fig:imaging}(b), respectively.

Momentum anti-correlation imaging is implemented analogously. Under the condition $\sigma_{p-}/\sigma_{p+}\gg1$, the normalized sum-coordinate intensity envelope can be approximated by $\delta(\mathbf{q}_+)$, confining the coincidence events to $\mathbf{q}_+=0$ and hence $\mathbf{q}_s=-\mathbf{q}_i$. Projection of the Fourier-plane JPD onto the minus- coordinate $\mathbf{q}_-$ yields the momentum-domain correlation image, whereas the complementary projection onto $\mathbf{q}_+$ reveals the narrow momentum anti-correlation peak. The momentum anti-correlation distribution and the reconstructed image are shown in Figs.~\ref{fig:imaging}(c) and \ref{fig:imaging}(d), respectively.

\section{Conclusion}
\label{sec:conclusion}

Here, we have demonstrated a bright source of high-dimensional spatially entangled photon pairs using a BBO crystal operated under beam-like Type-II phase-matching conditions. Compared with a conventional collinear Type-I SPDC source under identical pumping and acquisition conditions, our source exhibits $\sim$1.56- and $\sim$1.45-fold enhancements in the detected single-photon and coincidence biphoton count rates, respectively. Measurements in the position and momentum domains reveal strong position correlations and momentum anti-correlations.

Building on these spatial correlations, we further certified the dimensionality of the spatial entanglement in the generated biphoton state through correlation measurements in discretized transverse-position and transverse-momentum bases forming a pair of MUBs. Within a spatial dimension of $d=121$, the measured fidelity lower bound of $F_{\min}=0.5052$ certifies spatial entanglement in at least 62 dimensions. The position correlation and momentum anti-correlation were also exploited for correlation imaging, enabling object reconstruction in both spatial domains.

The demonstrated beam-like Type-II SPDC source combines enhanced brightness, strong spatial correlations, and high-dimensional entanglement within a simple experimental configuration. The source thus provides a versatile platform for high-dimensional quantum information processing, quantum imaging, quantum communication, and other photonic quantum technologies requiring bright, high-quality spatially entangled photon pairs.

\appendix*
\section{Fidelity-based certification of high-dimensional spatial entanglement}
\label{sec:supp}
Following the two-basis fidelity-certification method introduced in Refs.~\cite{Ndagano2020, Bavaresco2018}, we derive the fidelity lower bound used to certify the spatial entanglement dimensionality of the generated biphoton state.

The discrete position basis $\{|m\rangle\}_{m=0}^{d-1}$ and momentum basis $\{|u\rangle\}_{u=0}^{d-1}$ form a pair of mutually unbiased bases (MUBs), related through
\begin{equation}
\label{eq:mub}
\langle m|u\rangle=\frac{1}{\sqrt{d}}\exp\!\left(-\frac{2\pi i m u}{d}\right).
\end{equation}

The maximally entangled target state in the position basis is
\begin{equation}
|\Phi_d\rangle=\frac{1}{\sqrt{d}}\sum_{m=0}^{d-1}|m,m\rangle.
\end{equation}
Throughout this derivation, $\rho$ denotes the density operator restricted to and normalized within the selected $d\times d$ spatial-mode subspace, such that $\operatorname{Tr}(\rho)=1$.

The fidelity between the experimentally prepared state $\rho$ and the target state $|\Phi_d\rangle$ is defined as
\begin{equation}
\begin{split}
F(\rho,\Phi_d)
&=\mathrm{Tr}(\vert \Phi_d \rangle \langle \Phi_d \vert \rho)
=\langle\Phi_d|\rho|\Phi_d\rangle\\
&=\frac{1}{d}\sum_{m,n=0}^{d-1}\langle m,m|\rho|n,n\rangle.  
\end{split}
\end{equation}
Separating the diagonal and off-diagonal contributions gives
\begin{equation}
F(\rho,\Phi_d)=F_{\mathrm{diag}}+F_{\mathrm{coh}},
\end{equation}
where
\begin{align}
F_{\mathrm{diag}}&=\frac{1}{d}\sum_{m=0}^{d-1}\langle m,m|\rho|m,m\rangle,
\\
F_{\mathrm{coh}}&=\frac{1}{d}\sum_{\substack{m,n=0\\m\neq n}}^{d-1}\langle m,m|\rho|n,n\rangle.
\end{align}
The diagonal contribution $F_{\mathrm{diag}}$ is directly obtained from the diagonal elements of the JPD measured in the position basis.

To estimate $F_{\mathrm{coh}}$, we consider the
corresponding correlation measurement in the momentum basis.
The total probability under the momentum anti-correlated mode is 
\begin{equation}
\mathcal{T}=\sum_{u=0}^{d-1}\langle u,-u|\rho|u,-u\rangle.
\label{eq:momentum_projection}
\end{equation}
From Eq.~\eqref{eq:mub}, the two-photon momentum state can be expanded in the position basis as
\begin{equation}
\label{eq:momentum_expansion}
|u,-u\rangle=\frac{1}{d}\sum_{m,n=0}^{d-1}\exp\!\left[-\frac{2\pi i(m-n)u}{d}\right]|m,n\rangle.
\end{equation}
Substitution of Eq.~\eqref{eq:momentum_expansion} into Eq.~\eqref{eq:momentum_projection} yields
\begin{align}
\mathcal{T}&=\frac{1}{d^2}\sum_{u=0}^{d-1}\sum_{m,n,m',n'=0}^{d-1}\langle m,n|\rho|m',n'\rangle\nonumber\\
&\quad\times\exp\!\left[\frac{2\pi i\bigl(m-n-m'+n'\bigr)u}{d}\right]\nonumber\\
&=\frac{1}{d}\sum_{m,n,m',n'=0}^{d-1}\langle m,n|\rho|m',n'\rangle\,\delta_{\mathrm{mod}},
\label{eq:momentum_sum}
\end{align}
where
\begin{equation}
\delta_{\mathrm{mod}}\equiv\delta_{\left[(m-n)-(m'-n')\right]\bmod d,\,0}
\end{equation}
is the Kronecker delta enforcing
\begin{equation}
m-n\equiv m'-n'\pmod d.
\end{equation}

The quantity $\mathcal{T}$ can be decomposed as
\begin{equation}
\mathcal{T}=\mathcal{T}_0+\mathcal{T}_{\mathrm{coh}}+\mathcal{T}_{\mathrm{cross}},
\end{equation}
where
\begin{align}
\mathcal{T}_0&=\frac{1}{d}\sum_{m,n=0}^{d-1}\langle m,n|\rho|m,n\rangle=\frac{1}{d},\\
\mathcal{T}_{\mathrm{coh}}&=\frac{1}{d}\sum_{\substack{m\neq n}}^{d-1}\langle m,m|\rho|n,n\rangle=F_{\mathrm{coh}},\\
\mathcal{T}_{\mathrm{cross}}&=\frac{1}{d}\sum_{\substack{m\neq m',\,m\neq n\\n\neq n',\,n'\neq m'}}^{d-1}\langle m,n|\rho|m',n'\rangle\,\delta_{\mathrm{mod}}.
\end{align}

Positivity of the density operator implies
\begin{equation}
\left|\langle m,n|\rho|m',n'\rangle\right|\leq\sqrt{\langle m,n|\rho|m,n\rangle\langle m',n'|\rho|m',n'\rangle}.
\end{equation}
The cross contribution is therefore bounded by
\begin{align}
\left|\mathcal{T}_{\mathrm{cross}}\right|&\leq\frac{1}{d}\sum_{\substack{m\neq m',\,m\neq n\\n\neq n',\,n'\neq m'}}^{d-1}\sqrt{\langle m,n|\rho|m,n\rangle\langle m',n'|\rho|m',n'\rangle
}\nonumber\\
&\quad\times\delta_{\mathrm{mod}}\equiv C_{\mathrm{cross}}.
\end{align}

Since
\begin{equation}
F_{\mathrm{coh}}=\mathcal{T}-\mathcal{T}_0-\mathcal{T}_{\mathrm{cross}},
\end{equation}
the fidelity satisfies
\begin{align}
F(\rho,\Phi_d)
&\geq F_{\mathrm{diag}}+\mathcal{T}-\frac{1}{d}-\left|\mathcal{T}_{\mathrm{cross}}\right|\nonumber\\
&\geq F_{\mathrm{diag}}+\mathcal{T}-\frac{1}{d}-C_{\mathrm{cross}}\equiv F_{\min}.
\end{align}
Accordingly, the fidelity lower bound can be written explicitly as
\begin{equation}
\begin{split}
F_{\min} &= \frac{1}{d}\sum_{m=0}^{d-1} \langle m,m| \rho | m,m \rangle + \sum_{u=0}^{d-1} \langle u, -u | \rho | u, -u \rangle - \frac{1}{d}\\
 &- \frac{1}{d} \sum_{\substack{m\neq m',\, m\neq n\\ n'\neq n,\, m'\neq n'}} \sqrt{\langle m,n| \rho | m,n \rangle \langle m',n'| \rho | m',n' \rangle} \,\delta_{mod}.
\end{split}
\end{equation}

All quantities required to evaluate $F_{\min}$ are obtained from the JPDs measured in the position and momentum bases.  The corresponding density matrix elements are extracted according to
\begin{align}
\langle m,n| \rho | m,n \rangle &= \frac{\Gamma(m,n)}{\sum_{m,n=0}^{d-1} \Gamma(m,n)},\\
\langle u,v| \rho | u,v \rangle &= \frac{\Gamma(u,v)}{\sum_{u,v=0}^{d-1} \Gamma(u,v)}.
\end{align}
where $\Gamma_r$ and $\Gamma_q$ denote the experimentally reconstructed JPDs in the position and momentum domains, respectively. The resulting $F_{\min}$ provides a measurable lower bound on the fidelity to $|\Phi_d\rangle$. For this maximally entangled target state, $F_{\min}>k/d$ certifies a spatial entanglement dimensionality of at least $k+1$.

\begin{acknowledgments}
This work was supported by Quantum Science and Technology-National Science and Technology Major Project (Nos. 2024ZD0300900 and 2021ZD0301400); the National Natural Science Foundation of China (Nos. U25A20194, 12234009, 12274215, 12404382, 12427808, and 12625415); the Fundamental Research Funds for the Central Universities (2026300374); the Program for Innovative Talents and Entrepreneurs in Jiangsu; Key R\&D Program of Jiangsu Province (BE2023002); the Natural Science Foundation of Jiangsu Province (Nos. BK20233001).
\end{acknowledgments}

\section*{Conflicts of Interest}
The authors declare no conflicts of interest.

\section*{Data Availability}
The data that support the findings of this study are available from the corresponding author upon reasonable request.


\begin{thebibliography}{99}

\bibitem{Horodecki2009}
R. Horodecki, P. Horodecki, M. Horodecki, and K. Horodecki, Quantum entanglement, \href{https://doi.org/10.1103/RevModPhys.81.865}{Rev. Mod. Phys. \textbf{81}, 865 (2009)}.

\bibitem{Erhard2020}
M. Erhard, M. Krenn, and A. Zeilinger, Advances in high-dimensional quantum entanglement, \href{https://doi.org/10.1038/s42254-020-0193-5}{Nat. Rev. Phys. \textbf{2}, 365 (2020)}.

\bibitem{Zhang2024}
Z. Zhang, C. You, O. S. Maga{\~n}a-Loaiza, R. Fickler, R. de J. Le{\'o}n-Montiel, J. P. Torres, T. S. Humble, S. Liu, Y. Xia, and Q. Zhuang, Entanglement-based quantum information technology: A tutorial, \href{https://doi.org/10.1364/AOP.497143}{Adv. Opt. Photonics \textbf{16}, 60 (2024)}.

\bibitem{Cozzolino2019}
D. Cozzolino, B. Da Lio, D. Bacco, and L. K. Oxenl{\o}we, High-dimensional quantum communication: Benefits, progress, and future challenges, \href{https://doi.org/10.1002/qute.201900038}{Adv. Quantum Technol. \textbf{2}, 1900038 (2019)}.

\bibitem{Ecker2019}
S. Ecker, F. Bouchard, L. Bulla, F. Brandt, O. Kohout, F. Steinlechner, R. Fickler, M. Malik, Y. Guryanova, R. Ursin, and M. Huber, Overcoming noise in entanglement distribution, \href{https://doi.org/10.1103/PhysRevX.9.041042}{Phys. Rev. X \textbf{9}, 041042 (2019)}.

\bibitem{Luo2019}
Y.-H. Luo, H.-S. Zhong, M. Erhard, X.-L. Wang, L.-C. Peng, M. Krenn, X. Jiang, L. Li, N.-L. Liu, C.-Y. Lu, A. Zeilinger, and J.-W. Pan, Quantum teleportation in high dimensions, \href{https://doi.org/10.1103/PhysRevLett.123.070505}{Phys. Rev. Lett. \textbf{123}, 070505 (2019)}.

\bibitem{Moreau2019}
P.-A. Moreau, E. Toninelli, T. Gregory, and M. J. Padgett, Imaging with quantum states of light, \href{https://doi.org/10.1038/s42254-019-0056-0}{Nat. Rev. Phys. \textbf{1}, 367 (2019)}.

\bibitem{Defienne2024}
H. Defienne, W. P. Bowen, M. Chekhova, G. B. Lemos, D. Oron, S. Ramelow, N. Treps, and D. Faccio, Advances in quantum imaging, \href{https://doi.org/10.1038/s41566-024-01516-w}{Nat. Photonics \textbf{18}, 1024 (2024)}.

\bibitem{Ciampini2016}
M. A. Ciampini, N. Spagnolo, C. Vitelli, L. Pezz{\`e}, A. Smerzi, and F. Sciarrino, Quantum-enhanced multiparameter estimation in multiarm interferometers, \href{https://doi.org/10.1038/srep28881}{Sci. Rep. \textbf{6}, 28881 (2016)}.

\bibitem{Paesani2021}
S. Paesani, J. F. F. Bulmer, A. E. Jones, R. Santagati, and A. Laing, Scheme for universal high-dimensional quantum computation with linear optics, \href{https://doi.org/10.1103/PhysRevLett.126.230504}{Phys. Rev. Lett. \textbf{126}, 230504 (2021)}.

\bibitem{Chi2022}
Y. Chi, J. Huang, Z. Zhang, J. Mao, Z. Zhou, X. Chen, C. Zhai, J. Bao, T. Dai, H. Yuan, M. Zhang, D. Dai, B. Tang, Y. Yang, Z. Li, Y. Ding, L. K. Oxenl{\o}we, M. G. Thompson, J. L. O'Brien, Y. Li, Q. Gong, and J. Wang, A programmable qudit-based quantum processor, \href{https://doi.org/10.1038/s41467-022-28767-x}{Nat. Commun. \textbf{13}, 1166 (2022)}.

\bibitem{Chi2023}
Y. Chi, Y. Yu, Q. Gong, and J. Wang, High-dimensional quantum information processing on programmable integrated photonic chips, \href{https://doi.org/10.1007/s11432-022-3602-0}{Sci. China Inf. Sci. \textbf{66}, 180501 (2023)}.

\bibitem{Law2004}
C. K. Law and J. H. Eberly, Analysis and interpretation of high transverse entanglement in optical parametric down conversion, \href{https://doi.org/10.1103/PhysRevLett.92.127903}{Phys. Rev. Lett. \textbf{92}, 127903 (2004)}.

\bibitem{Howell2004}
J. C. Howell, R. S. Bennink, S. J. Bentley, and R. W. Boyd, Realization of the Einstein--Podolsky--Rosen paradox using momentum- and position-entangled photons from spontaneous parametric down conversion, \href{https://doi.org/10.1103/PhysRevLett.92.210403}{Phys. Rev. Lett. \textbf{92}, 210403 (2004)}.

\bibitem{Walborn2010}
S. P. Walborn, C. H. Monken, S. P{\'a}dua, and P. H. Souto Ribeiro, Spatial correlations in parametric down-conversion, \href{https://doi.org/10.1016/j.physrep.2010.06.003}{Phys. Rep. \textbf{495}, 87 (2010)}.

\bibitem{Schneeloch2016}
J. Schneeloch and J. C. Howell, Introduction to the transverse spatial correlations in spontaneous parametric down-conversion through the biphoton birth zone, \href{https://doi.org/10.1088/2040-8978/18/5/053501}{J. Opt. \textbf{18}, 053501 (2016)}.

\bibitem{Kwon2008}
O. Kwon, Y.-W. Cho, and Y.-H. Kim, Single-mode coupling efficiencies of type-II spontaneous parametric down-conversion: Collinear, noncollinear, and beamlike phase matching, \href{https://doi.org/10.1103/PhysRevA.78.053825}{Phys. Rev. A \textbf{78}, 053825 (2008)}.

\bibitem{Zhang2021}
C. Zhang, Y.-F. Huang, B.-H. Liu, C.-F. Li, and G.-C. Guo, Spontaneous parametric down-conversion sources for multiphoton experiments, \href{https://doi.org/10.1002/qute.202000132}{Adv. Quantum Technol. \textbf{4}, 2000132 (2021)}.

\bibitem{Takeuchi2001}
S. Takeuchi, Beamlike twin-photon generation by use of type-II parametric downconversion, \href{https://doi.org/10.1364/OL.26.000843}{Opt. Lett. \textbf{26}, 843 (2001)}.

\bibitem{Niu2008}
X.-L. Niu, Y.-F. Huang, G.-Y. Xiang, G.-C. Guo, and Z. Y. Ou, Beamlike high-brightness source of polarization-entangled photon pairs, \href{https://doi.org/10.1364/OL.33.000968}{Opt. Lett. \textbf{33}, 968 (2008)}.

\bibitem{Wang2016}
X.-L. Wang, L.-K. Chen, W. Li, H.-L. Huang, C. Liu, C. Chen, Y.-H. Luo, Z.-E. Su, D. Wu, Z.-D. Li, H. Lu, Y. Hu, X. Jiang, C.-Z. Peng, L. Li, N.-L. Liu, Y.-A. Chen, C.-Y. Lu, and J.-W. Pan, Experimental ten-photon entanglement, \href{https://doi.org/10.1103/PhysRevLett.117.210502}{Phys. Rev. Lett. \textbf{117}, 210502 (2016)}.

\bibitem{Liu2026}
Z.-F. Liu, Z.-C. Ren, P. Wan, W.-Z. Zhu, Z.-M. Cheng, J. Wang, Y.-P. Shi, H.-B. Xi, M. Huber, N. Friis, X. Gao, X.-L. Wang, and H.-T. Wang, Heralded high-dimensional photon--photon quantum gate, \href{https://doi.org/10.1038/s41566-026-01846-x}{Nat. Photonics \textbf{20}, 460 (2026)}.

\bibitem{Edgar2012}
M. P. Edgar, D. S. Tasca, F. Izdebski, R. E. Warburton, J. Leach, M. Agnew, G. S. Buller, R. W. Boyd, and M. J. Padgett, Imaging high-dimensional spatial entanglement with a camera, \href{https://doi.org/10.1038/ncomms1988}{Nat. Commun. \textbf{3}, 984 (2012)}.

\bibitem{Ndagano2020}
B. Ndagano, H. Defienne, A. Lyons, \textit{et al.}, Imaging and certifying high-dimensional entanglement with a single-photon avalanche diode camera, \href{https://doi.org/10.1038/s41534-020-00324-8}{npj Quantum Inf. \textbf{6}, 94 (2020)}.

\bibitem{Defienne2022}
H. Defienne, P. Cameron, B. Ndagano, A. Lyons, M. Reichert, J. Zhao, A. R. Harvey, E. Charbon, J. W. Fleischer, and D. Faccio, Pixel super-resolution with spatially entangled photons, \href{https://doi.org/10.1038/s41467-022-31052-6}{Nat. Commun. \textbf{13}, 3566 (2022)}.

\bibitem{Monken1998}
C. H. Monken, P. H. Souto Ribeiro, and S. P{\'a}dua, Transfer of angular spectrum and image formation in spontaneous parametric down-conversion, \href{https://doi.org/10.1103/PhysRevA.57.3123}{Phys. Rev. A \textbf{57}, 3123 (1998)}.

\bibitem{Defienne2018}
H. Defienne, M. Reichert, and J. W. Fleischer, General model of photon-pair detection with an image sensor, \href{https://doi.org/10.1103/PhysRevLett.120.203604}{Phys. Rev. Lett. \textbf{120}, 203604 (2018)}.

\bibitem{Bavaresco2018}
J. Bavaresco, N. Herrera Valencia, C. Kl{\"o}ckl, \textit{et al.}, Measurements in two bases are sufficient for certifying high-dimensional entanglement, \href{https://doi.org/10.1038/s41567-018-0203-z}{Nat. Phys. \textbf{14}, 1032 (2018)}.

\bibitem{Valencia2020}
N. Herrera Valencia, V. Srivastav, M. Pivoluska, \textit{et al.}, High-dimensional pixel entanglement: Efficient generation and certification, \href{https://doi.org/10.22331/q-2020-12-24-376}{Quantum \textbf{4}, 376 (2020)}.

\end{thebibliography}
\end{document}